# A REAL WORLD MECHANISM FOR TESTING SATISFIABILITY IN POLYNOMIAL TIME

Bernd R. Schuh

Dr. Bernd Schuh, Bernhardstraße 165, D-50968 Köln, Germany; bernd.schuh@netcologne.de

*Abstract*.
Whether the satisfiability of any formula F of propositional calculus can be determined in polynomial time is an open question. I propose a simple procedure based on some real world mechanisms to tackle this problem. The main result is the blueprint for a machine which is able to test any formula in conjunctive normal form (CNF) for satisfiability in linear time. The device uses light and some electrochemical properties to function. It adapts itself to the scope of the problem without growing exponentially in mass with the size of the formula. It requires infinite precision in its components instead.

a. *Basics.*

Introductions to the problem of satisfiability and its importance for complexity theory can be found in textbooks and reviews (see e.g. [1],[2]). In this section I will introduce the basic notations we will use in the following. They are simplified versions of a strictly algebraic approach outlined in [5], [6]. Consider n Boolean variables $B_n:=\{a_1,a_2,...a_n\}$, where n is arbitrary, and in practice very large. There are infinitely many ways to combine the $a_i$ by means of the logical operations "AND" ( x ), "OR" (+) and negation ( $\tilde{\phantom{a}}$ ), each way resulting in a formula F of propositional calculus. Formulas F over $B_n$ can be grouped into classes of logically equivalent formulas, however. That is, formulas F and F' belong to the same class iff their values under any truth assignment T: $B_n \rightarrow \{0,1\}$ are the same. Members of different classes are logically inequivalent, i.e. there is at least one truth assignment for which their values differ. This finite set of classes we call $V_n$. It has dimension $2^t$, where $t = 2^n$ is the number of truth assignments for $B_n$. As a convention, we number the t possible truth assignments $T_j$ (for fixed n) according to $T_j=j_{binary}$, where $j_{binary}$ represents the n-dimensional binary string which gives the truth values of the n elements of $B_n$. (see also the table in (2))

Each element of $V_n$ (except the class of tautologies) can be represented by a formula which we choose to be given in conjunctive normal form (CNF):

(1a) $\qquad C = C_1 \times C_2 \times ... \times C_m$

where each clause is a disjunction

(1b) $\qquad C_j = L_{j1} + L_{j2} + ... + L_{jk(j)}$

with literals $L_{ij} \in \{a_1, \tilde{a}_1, a_2, \tilde{a}_2, ... a_n, \tilde{a}_n\}$.

This representation is not unique but convenient. We will exploit the fact that there is a one-to-one correspondence between the elements of $V_n$ and the t - dimensional binary strings (C=lll0ll0..0l0l0ll0), each digit denoting the truth value of C belonging to the corresponding truth assignment of the $a_k$ i.e. $C_j = T_j(C)$ for the j-th component of C in its string representation. The numbering convention is illustrated in the following table for the formula $C = a_1 + \tilde{a}_3$ :

(2)

| # of assignment | a3 | a2 | a1 | C |
|---|---|---|---|---|
| 0 | 0 | 0 | 0 | 1 |
| 1 | 0 | 0 | 1 | 1 |
| 2 | 0 | 1 | 0 | 1 |
| 3 | 0 | 1 | 1 | 1 |
| 4 | 1 | 0 | 0 | 0 |
| 5 | 1 | 0 | 1 | 1 |
| 6 | 1 | 1 | 0 | 0 |
| 7 | 1 | 1 | 1 | 1 |

Thus, e.g., assignment number 6 means, since 6 = ll0 : $a_3$ is true, $a_2$ true, $a_1$ false. And for C we write in binary notation C= l0l0llll meaning that, e.g., C is not satisfiable for assignments number 4 and 6 (convention: read from right to left, starting with position number 0!), or explicitely for [$a_1$ false, $a_2$ false and $a_3$ true] or [$a_1$ false, $a_2$ true, $a_3$ true].

Also the basic variables $a_1, a_2,...$ represent elements of $V_n$ and thus each has a binary counterpart. E.g. for n=3, in the representation used above :

$a_1$ = l0l0l0l0;   $a_2$ = ll00ll00 ;   $a_3$ = llll0000

(see also the example (3).)The general formula for the $t=2^n$ $a_k$ being, $T_s(a_k)$ denoting the s digit of the binary expansion for $a_k$:

(3)     $T_r(a_k) = \Sigma_s \Sigma_l \ \delta(r, s+2^k l)$

where $\delta$ is the Kronecker $\delta$ and the s and l sums run from $2^{k-1}$ to $2^k-1$ and from 0 to $2^{n-k}-1$ respectively. Instead of using this formula to construct a set $B_n$ it is simpler to procede recursively. $B_{n-1}$ given, simply double the string of each $a_i$ with i<n, and add a new variable $a_n$ = ll..ll00..00 with $2^{(n-1)}$ zeros and ones respectively.

Next we note that negation is simply achieved by substituting zeros for ones and vice versa. For the $a_i$ this means reading the binary representation backwards (from left to right instead of right to left).   E.g.: $\tilde{a}_3$ = 0000llll. Of course, this (the "backward reading") does not hold for formulas in general, but only for the $a_i$, as defined by (3). Also operations + and x  can be performed digitwise by simple rules, i.e.:  let q + r = s, and $q_N q_{N-1}...q_1$ , $r_N...r_1$ etc. be their binary representations; then

(4a)          $s_j = q_j + r_j - q_j \cdot r_j$

(here + and – have their usual meaning of adding and subtracing numbers). Similarly for s= q x  r :

(4b)          $s_j = q_j \cdot r_j$.

Finally, one can utilize the fact, that the logical operation x can be traced back to + by negation. Explicitely we will use

(5) $\qquad \tilde{C} = \tilde{C}_1 + \tilde{C}_2 + ... + \tilde{C}_m$

b. *General setup of machines solving the problem.*

To formulate the task of a machine solving the SAT-problem we use a notation familiar from Turing machine approaches. Thus, we start with a finite alphabet $A = \{a_1, \tilde{a}_1,...a_n, \tilde{a}_n, [ , ] , € \}$ and define a set of allowed words W as follows

(6) $\qquad W = \{w \,/\, w = [\,L\,L\, ...\, L\,]...[\,L\, ...\, L\,]\,€\,\}$

Each L stands for any of the literals $L_{ij}$. The notation is an obvious transcription of (1): we skip the symbols + and x , [ and ] stand for the beginning and end of a clause, respectively, and we introduce € to signify the end of a sequence. Each CNF-formula with n variables can be expressed by a word of the form (6). The subset of W which corresponds to satisfiable formulas is usually called SAT. We take $w \,\varepsilon\, W$ as input of our machine M.

Having in mind that we deal with two operations + and x , and that our machine should be able to determine satisfiability we characterize the inner states q of M by three parameters, $q = (\mu, \nu, s)$.

Ideally, the machine does the following: on input $L_{ij}$ in state $(\mu,\nu,s)$ it preserves the input, goes into state $(\mu+1,\nu,s)$ und proceeds to the next symbol of the input, formally:

(7a) $\qquad L_{ij}/(\mu,\nu,s) \longrightarrow L_{ij}/(\mu+1,\nu,s)/\text{next}$

Whenever a new clause (in the input word w) starts, the machine reads [ . If the satisfiability parameter s=0 already, then clauses already done by the machine turned out to be unsatisfiable. Thus the whole formula is unsatisfiable, and we want M to stop:

(7b) $\qquad [\,/\,(\mu,\nu,1) \longrightarrow 0\,/\,(\mu,\nu,0)\,/\,\text{stop}$

If, however [ is read in the state $(\mu,\nu,s=1)$ M is to proceed.

(7c) $\qquad [\,/\,(\mu,\nu,1) \longrightarrow [\,/\,(\mu,\nu,1)\,/\,\text{next}$

The interesting step is when a clause in the input word is finished, then M gets ] as input. In this case we set up M like a nondeterministic Turing machine with a branching into two paths:

(7d) $\qquad ]\,/\,(\mu,\nu,s) \begin{array}{c} \nearrow 0\,/\,(0,\nu+1,0)\,/\,\text{next} \\ \searrow 1\,/\,(0,\nu+1,1)\,/\,\text{next} \end{array}$

Let us assume, the machine goes into state s=0. Then it sets 0 as output, clears the storage µ=0 and adds a finished clause to the existing ones (ν —> ν+1). If [ comes next, M will stop according to 7b). If € comes next, we make it stop:

(7e)   € / (µ,ν,0)  ⟶  0 / (µ,ν,0)  / stop

If M after a ] comes out with s=1, (7c) makes it proceed, until € is reached:

(7f)   € / (µ,ν,1)  ⟶  1 / (µ,ν,1)  / stop

The final output, s=0 or s=1 tells us whether the input w represents a satisfiable formula (s=1) or not; in other words: whether w $\varepsilon$ SAT or not. Thus M determines in finitely many steps whether a formula of the form (6) is satisfiable. And it clearly does so in "polynomial time", i.e. in p(|w|) steps, where p is a polynomial in the input word length. Up to now nothing more is achieved than a lengthy explanation of why SAT belongs to NP.

The crucial point is: What happens in the black box represented by equations (7), and especially equ. (7d)? Is it possible to set up a procedure that makes a determined and reliable choice in (7d)?
One answer to the second question is, of course, the following: we can always solve any SAT-formula, simply by trying out all possible assignments. And we can build such a program into the black box. But, in general such a program will need t trials, and nothing is gained. That is the heart of the "P=NP?"-problem: Is it possible to construct a universal machine, or to find a universal procedure that identifies each w $\varepsilon$

SAT in polynomial time. Universality here implies that the procedure be basically the same for each number n of basic variables, or any length of the basic alphabet A. In the following I will suggest a special purpose machine to "open the black box" in (7d). Special purpose machines can be constructed which utilize physical laws and/or technical and biochemical mechanisms of the real world to carry out many standard calculations in one step. Quantum computers or DNA-calculators may serve as examples, ([3],[4]). For the purpose of determining satisfiability much simpler devices suffice. I will call machines which act on the basic logical variables directly and perform many digital operations in one step "physical reasoning machines" (PRM).

c. *General considerations on PRMs for SAT.*

The basic operations we deal with, the logical functions + and x, are of binary nature, "true" and "false". SAT-formulas are formulated in terms of basic logical variables, the literals. They also come in a binary appearance as $a_i$ and $\tilde{a}_i$. Thus, it is not surprising that solving problems with n variables in a binary universe lets one end up with running times of the order $t=2^n$. Therefore, our aim is to construct a machine whose basic entities are not the binary digits zero (0) and one (I) but the literals themselves; additionally, the machine is to be capable of performing the logical operations on the level of the literals as a whole.

Of course, when modelling the literals as "real" entities their binary nature must be "built in", also with respect to the operations + and x. Thus one must start with some quality to be negated. There are many ways to do so. We can think of 0 and I as transistor out- and inputs (the common bits), for instance. The quality in this case being an electrical signal on the microvoltage scale. We can think of 0 and I as the beads of an abacus or the spots of a laser pointer being present or not, for instance.

The quality to be negated being material presence in the first case, darkness (or light) in the second.

*d. The Logical disk operator (LDO)*

In the following we choose transparency as the quality to be negated. Think of the $a_i$ as transparent disks, divided into $2^n$ fields, half of which are made black and thus intransparent. Think of the fields as numbered from 0 to $2^n-1$ and the black fields ordered in correspondence to the binary representation of the $a_i$ as given by (3). As examples we give here the three disks (n=3) $a_1$, $a_2$, $a_3$.

3a.                                                                 3b.

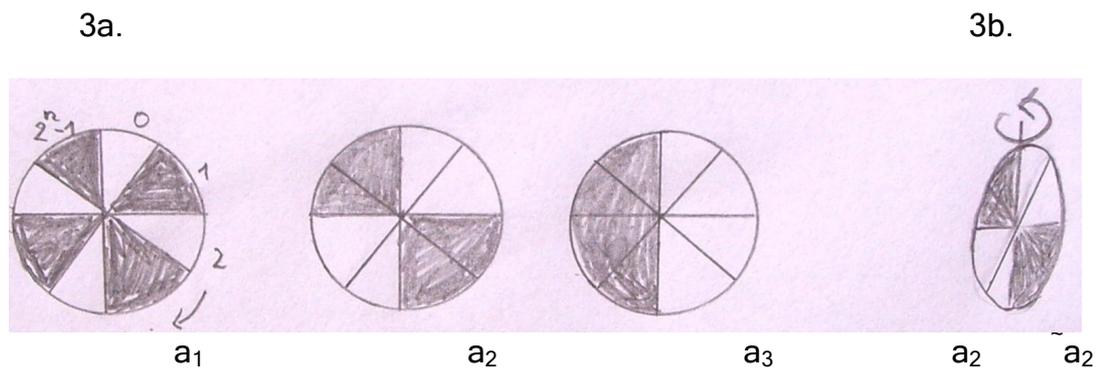

$a_1$            $a_2$            $a_3$            $a_2$    $\tilde{a}_2$

Figure 3 :   Representation of basic logical variables by disks

(As a convention we have chosen 12 o'clock as the fixed position from which the fields are numbered in clockwise direction.) How can we realize the basic logical operations with the disks? Note first, that rotating a disk by 180° around the axis separating the fields 0 to $2^{(n-1)}-1$ from the fields $2^{(n-1)}$ to $2^n-1$ is equivalent to negation of the $a_j$ represented by the disk; illustrated in Figure 3b for $a_2$.

Secondly, to do the logical OR (like $L_{ij} + L_{kl}$) operationally we only have to stack the corresponding disks on each other; the resulting image consisting of transparent and intransparent fields represents the outcome of the +-operation. Fig 2 illustrates the fact in the example $\tilde{a}_1 + a_2 + \tilde{a}_3$

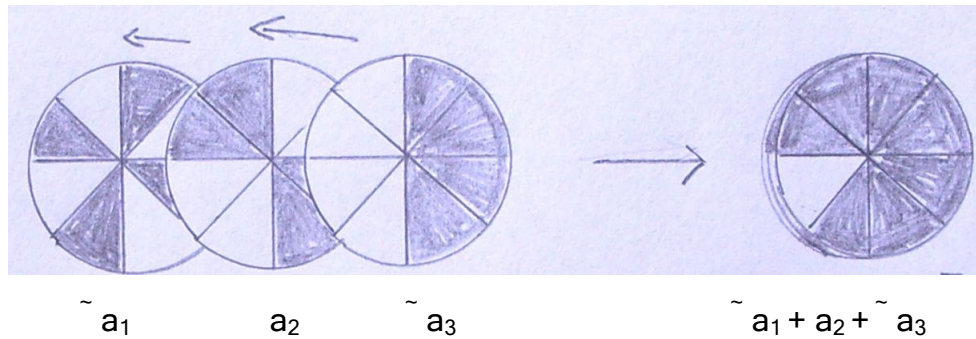

$\tilde{a}_1 \qquad a_2 \qquad \tilde{a}_3 \qquad\qquad \tilde{a}_1 + a_2 + \tilde{a}_3$

Figure 4 :  Logical "OR"   realized with disks

The result can be read off directly: the stack of disks is transparent on field number 5, so the formula of the example is false in one instant only, that is assignment 5, i.e. 5 = |0| = $a_3$ true, $a_2$ false, $a_1$ true.

Note that this rule of "addition" holds for all formulas, not only the a's.
Now think in clauses. So far, a complete clause $C_j$ corresponds to a stack of disks. If we illuminate this disk from below we see transparent and intransparent fields. This image corresponds to the evaluated clause $C_j$. The $C_j$, however, are to be multiplied (x). Since x can be transformed to + by negation (see equ. 5)) we can use the same procedure as before to generate ~C. Thus we need a mechanism to negate each $C_j$, or in terms of the coloured disks: we have to blacken the transparent fields and make the black ones transparent. For that purpose we assume to have additional transparent disks of a different nature; the disks of the second kind are made of a light sensitive material such that they can be blackened by light, similar to photosensitive paper. If we put such a disk on top of the stack it will get black at the transparent fields and stay transparent in the others. We now have a single disk representing $\tilde{C}_j$.

To complete the setup of the machine one proceeds as follows: in the working area for the clauses, WA1 say, the device generates the disks representing the $\tilde{C}_j$s. Those are "added" according to equ. (5), thus they are to be treated in the same way as the literals in WA1. So the machine stacks them in a second working area WA2. If all $\tilde{C}_j$s are collected on this stack its image under illumination shows $\tilde{C}$. The transparent fields of the whole stack in WA2 correspond to assignments for which $\tilde{C}$ is not satisfiable and thus C is satisfiable.

To set up a simple, one-step test of satisfiability we can put a photocell above the stack in WA2 and illuminate the stack from below. If the cell gives a signal C is satisfiable, otherwise not. It is convenient to assume that the test in WA2 is done with a light source different from the one in WA1, such that the photochemical process is not triggered in the transparent disks. Alternatively, one may fix the $\tilde{C}_j$ photochemically before moving them to WA2.

It is clear, also in this example of a PRM, that LDOs solve the satisfiability problem in polynomial time.

*e.*Description *in "Turing terms"*.

We choose the LDO of the foregoing section to make contact with the rules given in equ. (7). The model for a device obeying these rules consists of two working areas, WA1 and WA2, it furthermore has a storage area SA with a set of n disks representing the a's, as well as a basic supply storage BSS containing a sufficient number of transparent disks which change their transparency under (a certain) illumination. It furthermore has a light source under each working area, the one in WA1 triggering the photochemical process in the transparent disks, the other not. There is a photosensitive cell above WA2, which signals the parameter s. By $\mu$ and $\nu$ we number the disks in WA1 and WA2 respectively. If the photocell registers light it signals s=1, otherwise s=0.

The device starts out with empty stacks in WA1 and WA2, and s=1, i.e. q=(0,0,1). In this state the starting symbol [ changes nothing, the device reads the next letter of the input word, as stated by (7c). Given an $L_{ij}$ as input, i.e. an $a_j$ or $\tilde{a}_j$ the device (some robot in the black box) takes the corresponding disk from the storage, rotates it in case of an $\tilde{a}$, puts it on the stack in WA1, enlarges $\mu$ by 1 and reads the next symbol (see (7a)). As soon as ] is the input the device acts according to the following commands:

(i)   Take a transparent disk from BSS, put it on the stack in WA1, illuminate the whole stack from below.

(ii)  Once the transparent disk on top of the stack has been turned black in some fields remove it, and put it on the stack in WA2. Enlarge $\nu$ by 1.

(iii) Clear the stack in WA1, i.e. remove the disks from WA1, put them back into the storage. (this is not a necessary command for the functioning of the machine. It only makes the machine work with just one set of a-disks.)

(iv)  Illuminate the stack in WA2 from below and write the signal from the photocell above that stack as output, s.

(v)   Change the inner state s according to the outcome of that test.

These commands (i)-(v) carry out (7d). They guaranty that the choice in (7d) is completely deterministic and reflects the properties of the formula C under consideration, and nothing else.

If the photocell does not react, this means that the stack in WA2 is completely blackened, s=0. That makes the device stop on the next input letter which, from the grammar of allowed inputs, equ. (6), necessarily is [ or €.

As long as the photocell reacts there are still transparent fields in the stack in WA2. Consequently, ˜C is not satisfiable for assignments corresponding to these fields, and C is. Whether this property, s=1, persists we can only be certain of once all clauses have been worked out by the machine, i.e. the machine reaches the symbol € in state s=1, and the final output is 1.

In a straightforward modification the LDO can be made not only to test for satisfiability, but also to identify all assignments which satisfy the input formula. E.g. put as many photosensitive detectors above the stack in WA2 as there are fields on each disk. Correspondingly, one has $t=2^n$ parameters $s_j$. Then only those photocells will signal 1 which are located above a transparent region of the stack, all others 0. Obviously the problem arises of enumerating the assignments for which $s_i=1$ and one might easily run into an order $2^n$ problem.

So far we have discussed the operation of a single machine LDO(n) with a limited number of variables it can handle. This machine solves all problems with less than n variables. In order to construct one PRM that can solve SAT problems for any number of variables, we enrich the LDO by a set of instructions which enable it to enlarge itself: LDO(n) ➤ LDO(n+1).

### f. Self assembling logical disk operator (SALDO)

In addition to the working areas WA1 and WA2 there is an assembling area AA. It can be illuminated from below by the same light source as WA1. Furthermore, let $\alpha$ be the angle which determines the size of the one field in the disk with the largest number of fields ($a_1$ in our previous notation), i.e. $t \times \alpha_1 = 2\pi$. For reasons that become clear immideately we assume that the machine has two disks of this kind, representing the variable $a_1$. In a first step we copy the $a_1$ disk:

g0          Put the $a_1$ disk from SA to AA. Add a disk from BSS and illuminate.

As a result we get a $\sim a_1$ disk on top. It stays there.

g1          Rotate the disk on top counterclockwise by $\alpha_1/2$. Add a disk from BSS and illuminate.

g2          Rotate the disk on top counterclockwise by $\alpha_1$. Illuminate. Remove the top disk and fix it photochemically.

As a result of operations g0 - g2 the disk on top of AA has $2^{(n+1)}$ blackened fields of dimension $\alpha_1/2$ interchanging with as many transparent fields. Thus the newly generated disk represents a new $a_1$ disk. We could take care of this fact by renumbering all other $a_i$ ($a_1 \rightarrow a_2,...,a_n \rightarrow a_{n+1}$). But this is not necessary, as we will see soon. By fixing the result photochemically it is made resistant against the lightsource in WA1 (which it may encounter in the ongoing process of calculation). In order to have two copies of the new disk in store (as was assumed for the old $a_1$) the machine puts all disks in AA back into storage and repeats g0 with the new $a_1$. Thus the following step

g3          Remove the remaining disks in AA back to SA. Put $a_1$ back to AA and perform g0 once more. Fix the result and put both disks into storage SA.

completes the cycle of generating an additional a.

The process g0 - g2 is illustrated in figure 5 for the most simple case n=1.

g0

g1

g2

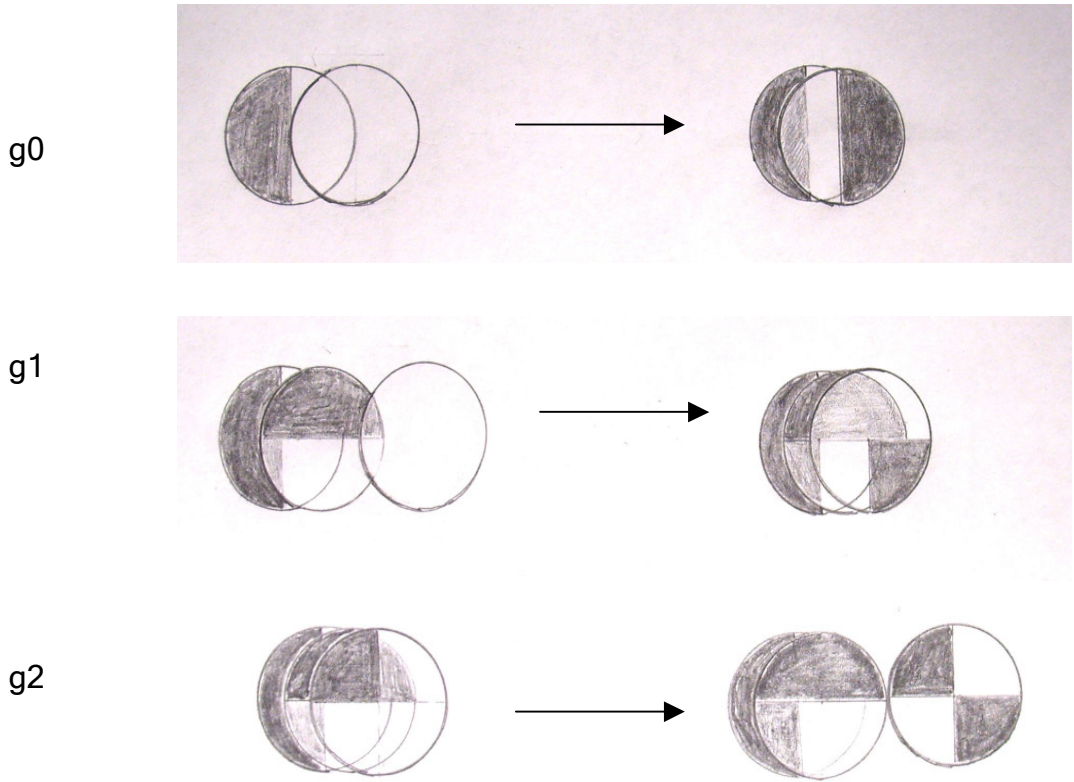

Fig. 5 : Illustration of procedure g0 to g2 for n=1

Essentially, the machine described in this section has enlarged its storage consisting of n distinct disks corresponding to n distinct $a_i$ by a new disk corresponding to a new $a_1$, i.e. the disk with the maximum number of fields in the machine. The assemblence is done in a number of steps that is independent of n. We can now define a self assembling PRM simply by adding the operational cycle g0 - g3 to the simplest LDO one can think of: LDO(1).

Before starting to operate it has only the two $a_1$ disks in SA and an (in principal) unlimited supply of transparent disks. But it assembles the missing ones "on demand": Whenever the machine encounters a literal in the input word, for which there is no disk, it assembles the "next" disk according to g0-g3, identifies it with the newly encountered literal, puts it (or its negative, according to the input word) in WA1 and proceeds. Note that for the self assembling device the numbering of the basic variables is not important. It defines its own variables according to the appearance of new literals in the input word. For the determination of satisfiability this implicit renumbering is not important. It becomes important only if one wants to determine the specific truth assignments for which the original formula is satisfied.

It is clear from the description of the self assembling procedure that the assembling cycle is done in linear time, independent of the number of fields on a disk or variables in storage. Therefore the SALDO tests any formula of propositional calculus given in CNF for satisfiability in time linear in the length of the input word.

*g. Conclusion.*

The main result is a construction plan for a logical calculator which determines the satisfiability of a Boolean formula F with n variables for any n (SALDO). It does so in linear time, working on F given as a string of variables and operational symbols. The device generates the necessary tools for calculating a formula F step by step on demand. It starts out with the bare light sources, the detection unit, a supply of transparent disks and a mask for the fabrication of the first variable. As soon as a new variable appears in the formula the machine assembles the corresponding disk (string, transparancy pattern, whatever you would like to call it) in one step. Since each new disk doubles the computing power of the machine as a whole its power increases exponentially with the length of the problem.Its material need does not grow exponentially, however, as one has in physical realizations like DNA-duplication machines ([3], [4]). In SALDO the material needs grow only linearly with the number of variables. The price to be payed for this is the decrease in area of the transparent pattern.

It is tempting to compare the SALDO with the purely mathematical notion of a Turing machine. All ingrediences of the SALDO resemble the parts of a Turing machine: the latter needs a reading head (one light source plus photocell), a writing head (light plus chemical properties of the disk material), and an infinite tape (unlimited supply of transparent disks). Even the starting disk (or a mask or blueprint to generate the first variable) is an implicit part of a Turing machine: its writing head must be able to distinguish the symbols 0 and 1.

Viewed as a "Gedanken" experiment the SALDO relies on two kinds of "infinities". The first is the unlimited supply of transparent disks. This assumption corresponds to the infinite tape in the analogy of a Turing machine. It may be accepted as a necessary but not very restrictive sine qua non. The second is the assumption of infinite precision, i.e. we assume that even infinitely small transparent fields can be clearly discerned from the non transparent sections and be adjusted correctly, when disks are put onto eachother  For a Gedanken experiment or a mathematical model this may be acceptable. For any implementation in a real world computer this property of the model is limiting, i.e. the assemblence procedure cannot be carried on ad infinitum with components of the real world. With growing n the fields on the disks get smaller and smaller and eventually one runs into precision problems with real world components, at latest when their size reaches the wavelength of the light used, probably earlier due to the photochemical processes involved.  A way out would be to enlarge the disks with growing n such that the area of each field was kept constant. That however implies an exponential growth in area and thus in the mass of the disks.

As a real world computer, the SALDO has further physical limitations. Its function relies on optical and chemical properties of matter,  and on some mechanical precision in handling the disks. Whether it is apt to perform actual calculations for problems of interest remains to be seen.

On the other hand, it appears legitimate to view the SALDO as a "Gedanken experiment" which can be carried through ("in mind") for any given degree of precision. As such it suggests that the long standing question "P=NP?" has a positive answer, because SAT is known to be NP-complete, and the SALDO solves SAT in polynomial time. Caution, however, appears to be in order. Because of its physical limitations the SALDO is not an ideal mathematical device like the Turing machine.Therefore its ability to check SAT in polynomial time can neither prove nor disprove P=NP, this question being defined in purely mathematical terms.